\documentstyle[preprint,epsfig,aps]{revtex}
\begin{document}
\draft

\preprint{\begin{tabular}{l}
\hbox to\hsize{\hfill {\bf (hep-ph/9905529)}}\\[-3mm]
\hbox to\hsize{\hfill YUMS 99-012}\\[-3mm]
\vspace{1.0cm}
\end{tabular} }


\title{ $b \to  s \gamma$ decays in the Left-Right Symmetric Model}
\author{C. S. Kim\thanks{kim@kimcs.yonsei.ac.kr,~~http://phya.yonsei.ac.kr/\~{}cskim/}
~ and ~ Yeong Gyun Kim\thanks{ygkim@cskim.yonsei.ac.kr} }
\address{Dept.~of Physics, Yonsei University,  Seoul 120-749, Korea}

\maketitle
\vspace{0.5cm}
\begin{center}
 (\today)
\end{center}

\begin{abstract}
\noindent
We consider $b \to  s \gamma$ decays in the Left-Right Symmetric Model.
Values of observables sensitive to chiral structure such as
the $\Lambda$ polarization in the $\Lambda_b \to  \Lambda \gamma$
decays and the mixing-induced CP asymmetries 
in the $B_{d,s} \to  M^0 \gamma$ decays can deviate
in the LRSM significantly from the SM values.
The combined analysis of $P_\Lambda$ and $A_{CP}$ as well as 
${\cal BR}(b \to  s \gamma)$ can be used to determine
the model parameters.
\end{abstract}


\newpage
\baselineskip .29in

\section{Introduction}

The Left-Right Symmetric Model (LRSM) \cite{LRSM} 
based upon the electroweak gauge group 
$SU(2)_L \times SU(2)_R \times U(1)$ represents well-known extensions of 
the Standard Model (SM), and can lead to interesting new physics effects 
in the $B$ system \cite{B1,B2}.
Due to the extended gauge structure there are both
new neutral and charged gauge bosons, $Z_R$ and $W_R$, as well as
a right-handed gauge coupling, $g_{_R}$.
The symmetry $SU(2)_L \times SU(2)_R \times U(1)$ can be broken to
$SU(2)_L \times U(1)$ by means of vacuum expectation values of 
doublet or triplet fields.
As for $SU(2)_L \times U(1)$ symmetry breaking, we assume that
it takes place when the scalar field $\Phi$ acquires
the complex vacuum expectation value
\begin{eqnarray}
<\Phi> =
\left( \begin{array}{cc} k_1 & 0 \\
                         0   & k_2~ e^{i \alpha} \\ 
       \end{array} \right) ~.
\end{eqnarray}
After symmetry breaking the charged $W_R$ mixes with $W_L$ of the SM to
form the mass eigenstates $W_{1,2}$ with eigenvalues $M_{1,2}$ 
and this mixing is described by two parameters; a real mixing angle $\zeta$ 
and a phase $\alpha$,
\begin{eqnarray}
\left( \begin{array}{c} W_1^+ \\ W_2^+ \\ \end{array} \right) =
\left( \begin{array}{cc} \cos ~\zeta & e^{-i \alpha} \sin ~\zeta \\
                        -\sin ~\zeta & e^{-i \alpha} \cos ~\zeta \\ 
       \end{array} \right)
\left( \begin{array}{c} W_L^+ \\ W_R^+ \\ \end{array} \right) ~.
\end{eqnarray}
The mixing angle $\zeta$ is small and can be expressed as
\begin{equation}
\zeta = {2 r \over 1 + r^2} {M_1^2 \over M_2^2} ,~~~~~(r \equiv k_2/k_1) ~.
\end{equation}
In this model the charged current interactions of the right-handed quarks 
are governed by a right-handed CKM matrix, $V_R$ , which , in principle, 
need not be related to its left-handed counterpart $V_L$.
Here we examine the possibility of using the rare decays $b \to  s \gamma$
as a new tool in exploring the parameter space of the LRSM.
We assume manifest left-right symmetry, that is $|V_R| = |V_L|$ 
and $\kappa \equiv g_{_R}/g_{_L} =1$.

The effective Hamiltonian of $b \to  s \gamma$ decay in the LRSM, after
ignoring $m_s$, is given by
\begin{equation}
H_{\rm eff} (b \to  s \gamma) = - { 4~G_F \over 
\sqrt{2}} V_{ts}^* V_{tb} \left[ C_{7L} O_{7L} + C_{7R} O_{7R} \right], 
\end{equation}
where
\begin{equation}
O_{7L}  = {e \over 16 \pi^2 }~m_b \bar{s}_L 
\sigma^{\mu\nu} b_R~F_{\mu\nu},~~~~~~
O_{7R}  = {e \over 16 \pi^2 }~m_b \bar{s}_R 
\sigma^{\mu\nu} b_L~F_{\mu\nu}.
\end{equation}
The magnetic moment operator coefficients 
are given by
\begin{eqnarray}
C_{7L} (m_b) &=& C_{7L}^{SM} (m_b) 
+ \zeta {m_t \over m_b} {V_R^{tb} \over V_L^{tb}} e^{i \alpha} 
\left[~\eta^{16/23} \tilde{F}(x_t) 
+ {8 \over 3} (\eta^{14/23} -\eta^{16/23}) \tilde{G} (x_t) \right] \nonumber
\\
&+& {2 r (1+r^2) \over (1-r^2)^2}
{m_t \over m_b} {V_R^{tb} \over V_L^{tb}} e^{i \alpha} 
\left[~\eta^{16/23} \tilde{F}_H (y_t) 
+ {8 \over 3} (\eta^{14/23} -\eta^{16/23}) \tilde{G}_H (y_t) \right],
\\
C_{7R} (m_b) &=& 
\zeta {m_t \over m_b} 
\left( {V_R^{ts} \over V_L^{ts}} \right)^* e^{-i \alpha} 
\left[~\eta^{16/23} \tilde{F}(x_t) 
+ {8 \over 3} (\eta^{14/23} -\eta^{16/23}) \tilde{G} (x_t) \right] \nonumber
\\
&+& {2 r (1+r^2) \over (1-r^2)^2} {m_t \over m_b} 
\left( {V_R^{ts} \over V_L^{ts}} \right)^* e^{-i \alpha} 
\left[~\eta^{16/23} \tilde{F}_H (y_t) 
+ {8 \over 3} (\eta^{14/23} -\eta^{16/23}) \tilde{G}_H (y_t) \right],
\end{eqnarray}
where
\begin{eqnarray}
&C&_{7L}^{SM} (m_b) = \eta^{16/23} F(x_t) 
+ {8 \over 3} (\eta^{14/23} - \eta^{16/23}) G(x_t) +
~\sum~ h_i \eta^{p_i} ~,
\end{eqnarray}
with $\eta = \alpha_s(M_1)/\alpha_s(m_b)$, $x_t = (m_t/M_1)^2$
and $y_t = (m_t/M_H)^2$, where $M_H$ is the mass of the charged physical
scalars.
The various functions of $x_t$, $y_t$, and
the coefficients $h_i$, and powers $p_i$ 
can be founded in the Ref. \cite{B2}.

\section{ Observables sensitive to chiral structure }

\subsection{Branching fraction of inclusive decay ${\cal BR}(b \to  s \gamma$) }

\noindent
The decay rate for inclusive $b \to  s \gamma$ decay is given by
\begin{equation}
\Gamma (b \to  s \gamma) = { G_F^2 m_b^5 \over 32 \pi^4 } \alpha_{em}
|V_{ts}^* V_{tb}|^2 \left( |C_{7L} (m_b)|^2 + |C_{7R} (m_b)|^2 \right) .
\end{equation}
It is common practice to normalize this radiative partial width to
the semileptonic rate
\begin{equation}
\Gamma (b \to  c e \bar{\nu}) =
{ G_F^2 m_b^5 \over 192 \pi^3 } |V_{cb}|^2 f({m_c \over m_b}) 
\left[ 1 - {2 \over 3 \pi} \alpha_s (m_b) g({m_c \over m_b}) \right] ,
\end{equation}
where $f(x)=1-8 x^2-24 x^4 \ln x +8 x^6-x^8$ represents a phase space factor, 
and the function $g(x)$ encodes next-to-leading order strong interaction 
effects \cite{cskim}.
In terms of the ratio $R$, 
\begin{equation}
R \equiv {\Gamma (b \to  s \gamma) 
\over \Gamma(b \to  c e \bar{\nu})} =
{6 \over \pi} {|V_{ts}^* V_{tb}|^2 \over |V_{cb}|^2}
{\alpha_{em} \over f({m_c \over m_b})}
{ |C_{7L} (m_b)|^2 + |C_{7R} (m_b)|^2 \over 
1 - {2 \over 3 \pi} \alpha_s (m_b) g({m_c \over m_b}) } ,
\end{equation}
the $b \to  s \gamma$ branching fraction is obtained by
\begin{equation}
{\cal BR}(b \to  s \gamma) = {\cal BR}(b \to  c e \bar{\nu}) \times R
\simeq {\cal BR}(B \to X_c l \nu)_{\rm exp.} \times R
\sim (0.105) \times R .
\end{equation}
In Eqs. (8,9), we neglected the $1/m_b^2$ corrections.
For ${\cal BR}(b \to  s \gamma)$, 
we also use the present experimental value \cite{CLEO} of the branching 
fraction for $B \to  X_s \gamma$ decay, 
\begin{equation}
{\cal BR}(B \to  X_s \gamma) = 
(3.15 \pm  0.35_{\rm stat} \pm 0.32_{\rm syst} \pm 0.26_{\rm model}) \times 10^{-4} .
\end{equation}

\subsection{$\Lambda$ Polarization 
in  $\Lambda_b \to  \Lambda \gamma$ decay}

\noindent
One way to access the chiral structure is to consider the decay of baryons. 
{}From the experimental side the decay $\Lambda_b \to  \Lambda \gamma$ 
is a good candidate, 
since the subsequent $\Lambda$ decay $\Lambda \to  p \pi$
is self analyzing \cite{mannel}.
The expected branching ratio is of order $10^{-5}$ and should be measurable at
future hadronic $B$ factories, HERA-B, BTeV and LHC-B.
The chiral structure can be studied by measuring the polarization of $\Lambda$,
via the angular distribution,
\begin{eqnarray}
{1 \over \Gamma} {d \Gamma \over d \cos \theta} =
{1 \over 2} (1 + P_{\Lambda} \cos \theta) ,
\end{eqnarray}
where
\begin{eqnarray}
P_{\Lambda} = 
{ |C_{7L} (m_b)|^2 - |C_{7R} (m_b)|^2 
\over |C_{7L} (m_b)|^2 + |C_{7R} (m_b)|^2 } ,
\end{eqnarray} 
and $\theta$ is the angle between the direction of the momentum of $\Lambda$
in the rest frame of $\Lambda_b$ and the direction of the $\Lambda$ polarization
in the $\Lambda$ rest frame.

\subsection{Mixing-induced CP Asymmetry in $B_{d,s} \to  M^0 \gamma$ decays}

\noindent
Next, we consider the mixing-induced CP asymmetry for 
$B_{d,s} \to  M^0 \gamma$ decays \cite{atwood}.
Here $M^0$ is any hadronic self-conjugate state, with CP eigenvalue $\xi = \pm 1$.
The decay amplitudes are denoted by
\begin{eqnarray}
A({\bar B}_{d,s} \to  M^0 \gamma_{_L}) &=& A~ \cos~ \psi~ e^{i \phi_{_L}}, \nonumber \\
A({\bar B}_{d,s} \to  M^0 \gamma_{_R}) &=& A~ \sin~ \psi~ e^{i \phi_{_R}}, \nonumber \\
A(B_{d,s} \to  M^0 \gamma_{_R}) &=& \xi~ A~ \cos~ \psi~ e^{-i \phi_{_L}}, \nonumber \\
A(B_{d,s} \to  M^0 \gamma_{_L}) &=& \xi~ A~ \sin~ \psi~ e^{-i \phi_{_R}}.
\end{eqnarray}
Here the parameter $\psi$ gives the relative amount of 
left-polarized photons compared to right-polarized photons 
in ${\bar B}_{d,s}$ decays, and $\phi_{L,R}$ are CP-odd weak phases.
Using the time dependent rates $\Gamma(t)$ and $\bar{\Gamma} (t)$ for
$B_{d,s} \to  M^0 \gamma$ and ${\bar B}_{d,s} \to  M^0 \gamma$ respectively,
one finds a time-dependent CP asymmetry 
\begin{eqnarray}
A (t) = {\Gamma(t) - \bar{\Gamma} (t) \over \Gamma(t) + \bar{\Gamma} (t) }
= \xi~ A_{CP}~ \sin(\Delta m t) ,
\end{eqnarray}
where
\begin{eqnarray}
A_{CP} \equiv \sin(2 \psi) \sin(\phi_{_M} - \phi_{_L} - \phi_{_R}) ,
\end{eqnarray}
and $\Delta m$ and $\phi_{_M}$ are the mass difference and 
phase in the $B_{d,s} - {\bar B}_{d,s}$ mixing amplitude.

In terms of $C_{7L(R)}$, the $\psi$ and $\phi_{L(R)}$ are given by
\begin{eqnarray}
\sin(2 \psi) = {2 |C_{7L}(m_b)~C_{7R}(m_b)| 
\over |C_{7L}(m_b)|^2 + |C_{7R}(m_b)|^2} ~,
\end{eqnarray}
\begin{eqnarray}
\phi_{_L} = \sin^{-1} \left( Im(C_{7L}(m_b)) \over |C_{7L}(m_b)| \right),~~~~~~
\phi_{_R} = \sin^{-1} \left( Im(C_{7R}(m_b)) \over |C_{7R}(m_b)| \right).
\end{eqnarray}
The phase of $B_{d,s}-{\bar B}_{d,s}$ mixing can be also affected 
by new LRSM contributions \cite{barenboim}, and is given by 
$\phi_{_M} = \phi_{_M}^{SM} + \delta_{_M}$ where 
\begin{eqnarray}
\delta_{_M} = \tan^{-1} \left({h~\sin \sigma \over 1+h~\cos \sigma} \right).
\end{eqnarray} 
Here $h = |M_{12}^{LR}|/|M_{12}^{SM}|$ measures the relative size
of the left-right contribution to the non-diagonal element $M_{12}$
and can be written as
\begin{eqnarray}
h = F(M_2) \left( {1.6~{\rm TeV} \over M_2}\right)^2 + 
    \left( {12~{\rm TeV} \over M_H} \right)^2,
\end{eqnarray} 
where $F(M_2)$ is a complicated function of the $M_2$.
The phase $\sigma$ can be expressed as
\begin{eqnarray}
e^{i \sigma} = 
- {V_R^{td*}~ V_R^{tb} \over V_L^{td*}~ V_L^{tb}}~,~~~~
e^{i \sigma} = 
- {V_R^{ts*}~ V_R^{tb} \over V_L^{ts*}~ V_L^{tb}}~,
\end{eqnarray}
for $B_d$ and $B_s$ systems, respectively. And
$\phi_{_M}^{SM} = 2 \beta$ and $\phi_{_M}^{SM} = 0$ for 
$B_d$ and $B_s$ systems, respectively 
(where $-\beta$ is the phase of $V_{td}$ in the standard convention).

\section{Combined Analysis}

\noindent
In this Section, we perform the combined analysis of three observables,
${\cal BR} (b \to s \gamma)$, $P_{\Lambda}$ and $A_{CP}$. 
Fig. 1 is the contour plot for ${\cal BR}(b \to  s \gamma)$ and $P_\Lambda$
on the ($|C_{7L}(m_b)|$ ,$~~|C_{7R}(m_b)|$) plane.
Two solid curves indicate the $1 \sigma$ range of 
the measured values of inclusive ${\cal BR}(b \to  s \gamma)$.
Three dashed lines correspond to 
three different values of $P_\Lambda$, as indicated in the figure.
{}From the measurements of ${\cal BR}(b \to  s \gamma)$ and $P_\Lambda$,
one can determine the magnitudes $|C_{7L}(m_b)|$ and $|C_{7R}(m_b)|$ separately.
And the measurement of $A_{CP}$ would give some informations on the phases of
$C_{7L}(m_b)$ and $C_{7R}(m_b)$.

\subsection{Simple Case}

\noindent
First, we consider a simple case. We assume $V_L = V_R$ and ignore
the contributions from $W_2^\pm$ and charged physical scalars.
Then only two new physics parameters, $\zeta$ and $\alpha$, remain.
To illustrate the usefulness of $P_\Lambda$ measurements,
let's consider $\alpha = 0$ case.
Fig. 2(a) shows the dependence of inclusive ${\cal BR}(b \to  s \gamma)$ on
the mixing angle $\zeta$ in this case. 
Two horizontal dashed lines indicate the $1 \sigma$ range of 
the present measured values of inclusive ${\cal BR}(b \to  s \gamma)$.
It is clear from the figure that 
the SM result is essentially obtained when $\zeta=0$, and also
that a conspiratorial solution occurs when $\zeta \sim -0.01$.
These two cases are indistinguishable, and even independent of any further 
improvements in the measurement of the inclusive ${\cal BR}(b \to  s \gamma)$.
However, if $P_\Lambda$ in $\Lambda_b \to  \Lambda \gamma$ decays
is measured in addition, these two solutions are definitely distinguishable as 
indicated in Fig. 2(b), 
which shows the dependence of $P_\Lambda$ on the mixing angle $\zeta$.
The $\zeta \sim 0$ case corresponds to $P_\Lambda \sim +1$, and
the $\zeta \sim -0.01$ case corresponds to $P_\Lambda \sim -1$.

When we vary the phase $\alpha$ between 0 and $\pi$ radian, 
$P_\Lambda$ can have all the possible values from $+1$ to $-1$,
while satisfying the inclusive ${\cal BR}(b \to  s \gamma)$ constraints.
Figs. 3(a) and 3(b) show the dependence of $P_\Lambda$ on 
$\zeta$ and $\alpha$ respectively, where we impose 
the present experimental ${\cal BR}(B \to  X_s \gamma)$ constraints. 
The larger magnitudes of $\zeta$ gives larger deviations of 
$P_\Lambda$ from the SM expectation, $P_\Lambda (SM) = +1$.
Because the measurements of ${\cal BR}(b \to  s \gamma)$ and $P_\Lambda$
determine only the magnitudes of $C_{7L} (m_b)$ and $C_{7R} (m_b)$,
the $\zeta$ can be determined up to the sign ambiguity.

Next, we consider $A_{CP}$ in the radiative $B_{d,s}$ decays,
$B_{d,s} \to  M^0 + \gamma$, {\it eg.}, $B_{d} \to K^* + \gamma$ and
$B_{s} \to  \phi + \gamma$.
The dependences of $A_{CP}$ on the $\zeta$ and $\alpha$ is shown
in  Figs. 4(a) and 4(b), respectively,
for $B_d \to  M^0 \gamma$ decay. And in Figs. 4(c) and 4(d)  we show 
the dependence of $A_{CP}$ on the $\zeta$ and $\alpha$
for $B_s \to  M^0 \gamma$ decay.
Here we impose 
the present experimental inclusive ${\cal BR}(B \to  X_s \gamma)$ 
constraints \cite{CLEO}.
For numerical value of $\beta$, we use the central value of 
the recent CDF measurement \cite{CDF} of
$\sin~2\beta$ from $B_d \to J/\psi + K_s$,
\begin{equation}
\sin~2\beta = 0.79^{+0.41}_{-0.44} .
\end{equation}
It is clear from the figures that  $A_{CP}$ can have rather large
values between $-20 \%$ and $90 \%$ for $B_d \to  M^0 \gamma$ decay, and
up to $\pm 60 \%$ for $B_s \to  M^0 \gamma$ decay, 
while the SM expectation values
$A_{CP} (SM)$ are almost zero. 
We can see that different sign of $\zeta$ with same magnitude
correspond to different values of $A_{CP}$.
Therefore, the sign ambiguity of $\zeta$ determined from $P_\Lambda$
measurements can be resolved by measuring $A_{CP}$.
Moreover, $P_\Lambda$ and $A_{CP}$ have definite correlations,
as shown in Figs. 5(a) and (b) for $B_d \to  M^0 \gamma$ and
for $B_s \to M^0 \gamma$, respectively.
Any deviations from these correlations
would indicate the failure of the manifest left-right symmetric 
scenario which we assume in this subsection.

\subsection{General Case}

\noindent
Now we consider more general case. We assume that the elements of $V_R$ have
arbitrary phase. We also consider the contributions from $W_2^\pm$ and 
charged physical scalars. 
In this case, $C_{7L}(m_b)$ 
depends on the parameters; $r, \omega_1, M_2$ and
$M_H$. And $C_{7R}(m_b)$ depends on  $r, \omega_2, M_2$ and $M_H$. 
The parameters $\omega_1$ and $\omega_2$ are defined as
\begin{eqnarray}
e^{i \omega_1} \equiv {V_R^{tb} \over V_L^{tb}} e^{i \alpha},~~~~~~
e^{i \omega_2} \equiv {V_R^{ts} \over V_L^{ts}} e^{i \alpha}.
\end{eqnarray}
For further numerical calculations, we fix $M_2 = 1.6$~TeV and $M_H = 12$~TeV.

While the magnitude of $C_{7L}(m_b)$ depends on the $r$ and $\omega_1$,
the magnitude of $C_{7R}(m_b)$ only on the $r$ but not on the $\omega_2$.
Therefore, $P_\Lambda$ depends on the $r$ and $\omega_1$ 
but not on the $\omega_2$.
The dependences of $P_\Lambda$ on the $r$ and $\omega_1$ are shown
in Figs. 6(a) and 6(b), respectively. 
For large $r$, the value of $P_\Lambda$ can be largely deviated 
from the SM value due to the large contributions from charged physical scalars
even though $\zeta$ is small. From the measurement of $P_\Lambda$ we can
determine the values of $r$ and $\omega_1$ (up to discrete ambiguity)
for given values of $M_2$ and $M_H$.

In $B_d \to M^0 \gamma$ decays, $A_{CP}$ has additional dependences
on another new phase $\omega_3$ and also on phase $\beta$ through
$\phi_M$, the phase of $B_{d}-{\bar B}_{d}$ mixing. The phase $\omega_3$
is defined by
\begin{eqnarray}
e^{i \omega_3} \equiv {V_R^{td} \over V_L^{td}} e^{i \alpha}~.
\end{eqnarray}
The dependence of $A_{CP}$ on $\omega_2$ appear only through $\phi_R$
in this case. And the phase of $B_{d}-{\bar B}_{d}$ mixing, $\phi_M$
would be determined independently 
from the measurement of $A_{J/\Psi K_s}$,
{\it i.e.} the mixing induced CP aymmetry 
in the $B_d \to J/\Psi + K_s$ decays \cite{CDF},
\begin{eqnarray}
A_{J/\Psi K_s} = \sin(\phi_M)~.
\end{eqnarray}
Therefore, in addition to $P_\Lambda$, for the $B_d$ system 
the value of $\omega_2$ can be determined up to discrete ambiguity
for given values of $M_2$ and $M_H$ from the measurement of $A_{CP}$.
In  $B_s \to M^0 \gamma$ decays, $A_{CP}$ has dependence on
the $\omega_2$ through $\phi_R$ and also on $\phi_M$. 
As can be seen from Eq. (18), $A_{CP}$ can have any values
between $-\sin(2 \Psi)$ and $+\sin(2 \Psi)$ depending on the $\omega_2$.
The dependences of $\sin(2 \Psi)$, the maximum value of $A_{CP}$,
on the $r$ and $\omega_1$ are shown in
the Figs. 7(a) and 7(b), respectively, for $B_{d,s} \to M^0 \gamma$ decays. 
It is clear that
the values of $A_{CP}$ can be largely deviated from the SM prediction. 
From the measurement of $A_{CP}$ in addition to $P_\Lambda$,
the value of $\omega_2$ can be also determined up to discrete ambiguity
for given values of $M_2$ and $M_H$.

\vspace{0.5cm}
To summarize, in this paper we considered the radiative $B$ hadron decay
in the Left-Right Symmetric Model (LRSM).
Values of observables sensitive to chiral structure such as
the $\Lambda$ polarization in the $\Lambda_b \to  \Lambda \gamma$
decays and the mixing-induced CP asymmetries 
in the $B_{d,s} \to  M^0 \gamma$ decays can deviate
in the LRSM significantly from the SM values.
The combined analysis of $P_\Lambda$ and $A_{CP}$ as well as 
${\cal BR}(b \to  s \gamma)$ can be used to determine
the model parameters.
{}From the correlations between $P_\Lambda$ and $A_{CP}$,
the validity of the manifest left-right symmetry scenario can also
be tested.

\acknowledgements

\noindent
We thank G. Cvetic and T. Morozumi for careful reading of the manuscript and their
valuable comments.
C.S.K. wishes to acknowledge the financial support of 1997-sughak program of 
Korean Research Foundation, Project No. 1997-011-D00015.
The work of Y.G.K. was supported by KOSEF Postdoctoral Program.


%
%
\begin{figure}[ht]
\hspace*{-1.0 truein}
\psfig{figure=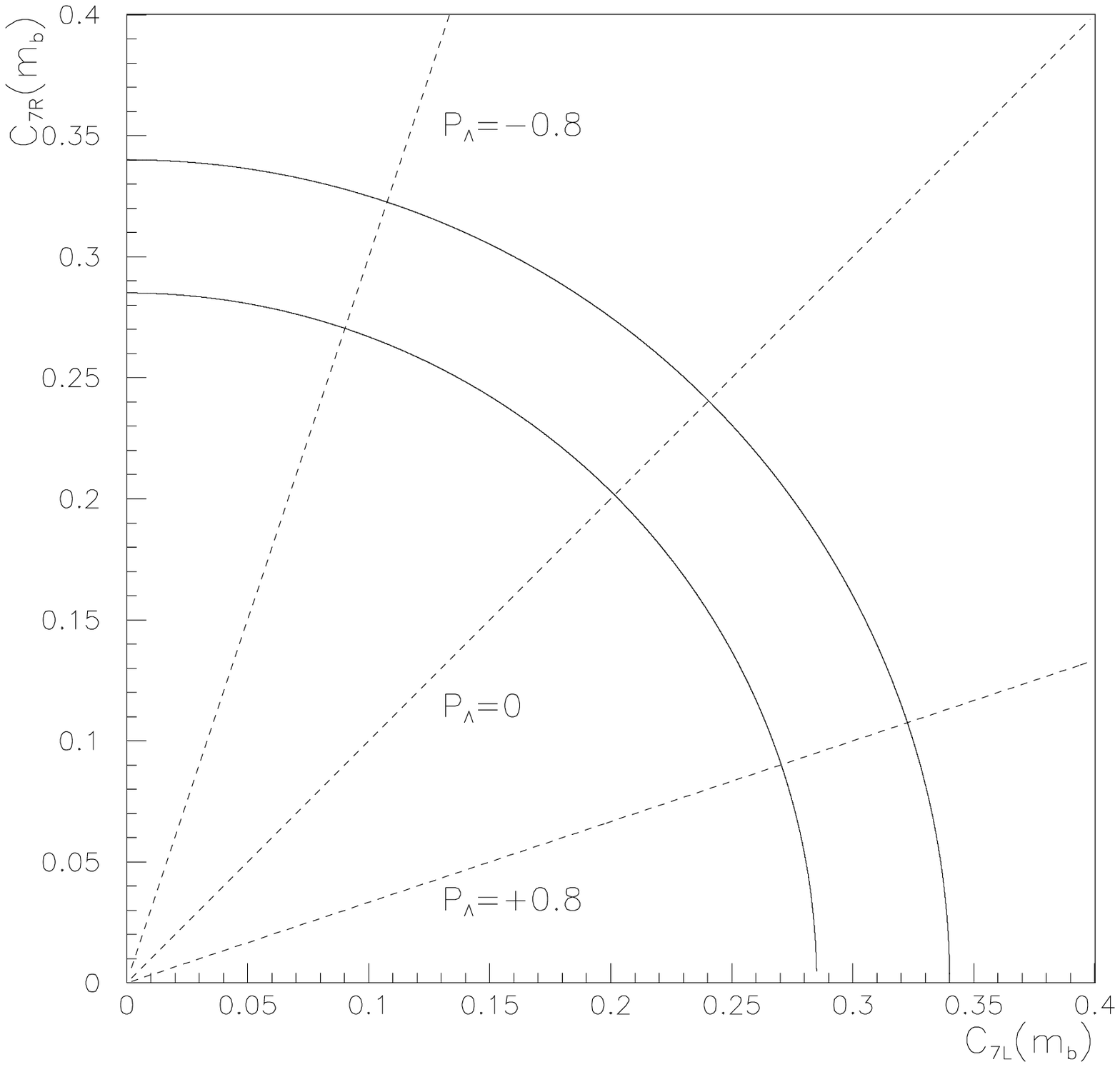}
\caption{ Contour plots for the inclusive  ${\cal BR}(b \to  s \gamma)$ and $P_\Lambda$.
Two solid curves indicate the $1 \sigma$ range of 
the present measured values of inclusive ${\cal BR}(B \to  X_s \gamma)$.}
\label{fig1}
\end{figure}

\begin{figure}[ht]
\hspace*{-1.0 truein}
\psfig{figure=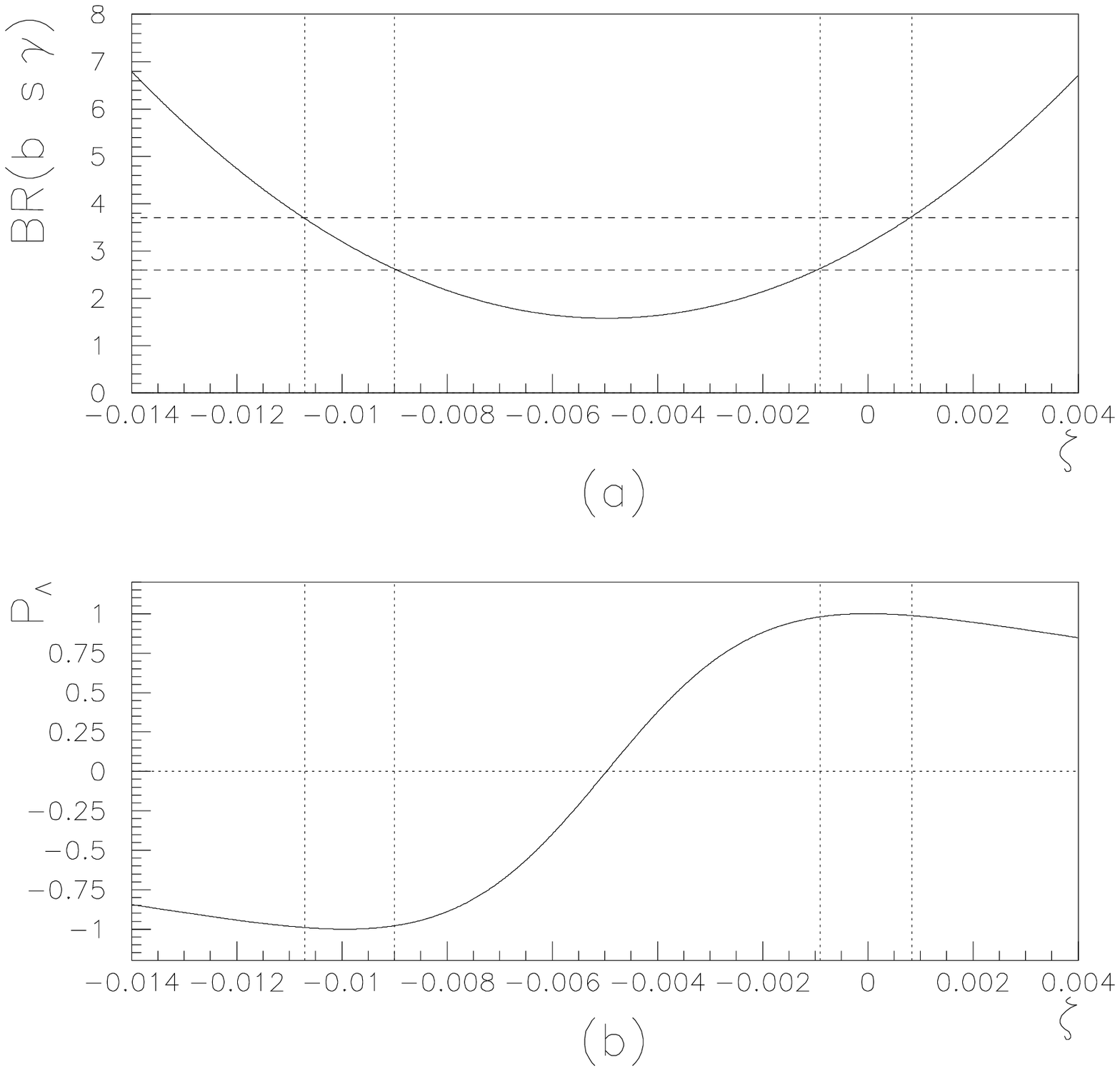}
\caption{(a) Dependence of inclusive ${\cal BR}(b \to  s \gamma)$ 
on mixing angle $\zeta$. Two horizontal dashed lines indicate the $1 \sigma$ range of 
the present measured values of inclusive ${\cal BR}(B \to  X_s \gamma)$.
(b) Dependence of $P_\Lambda$ on mixing angle
$\zeta$. In both cases we fix $\alpha = 0$.
}
\label{fig2}
\end{figure}

\begin{figure}[ht]
\hspace*{-1.0 truein}
\psfig{figure=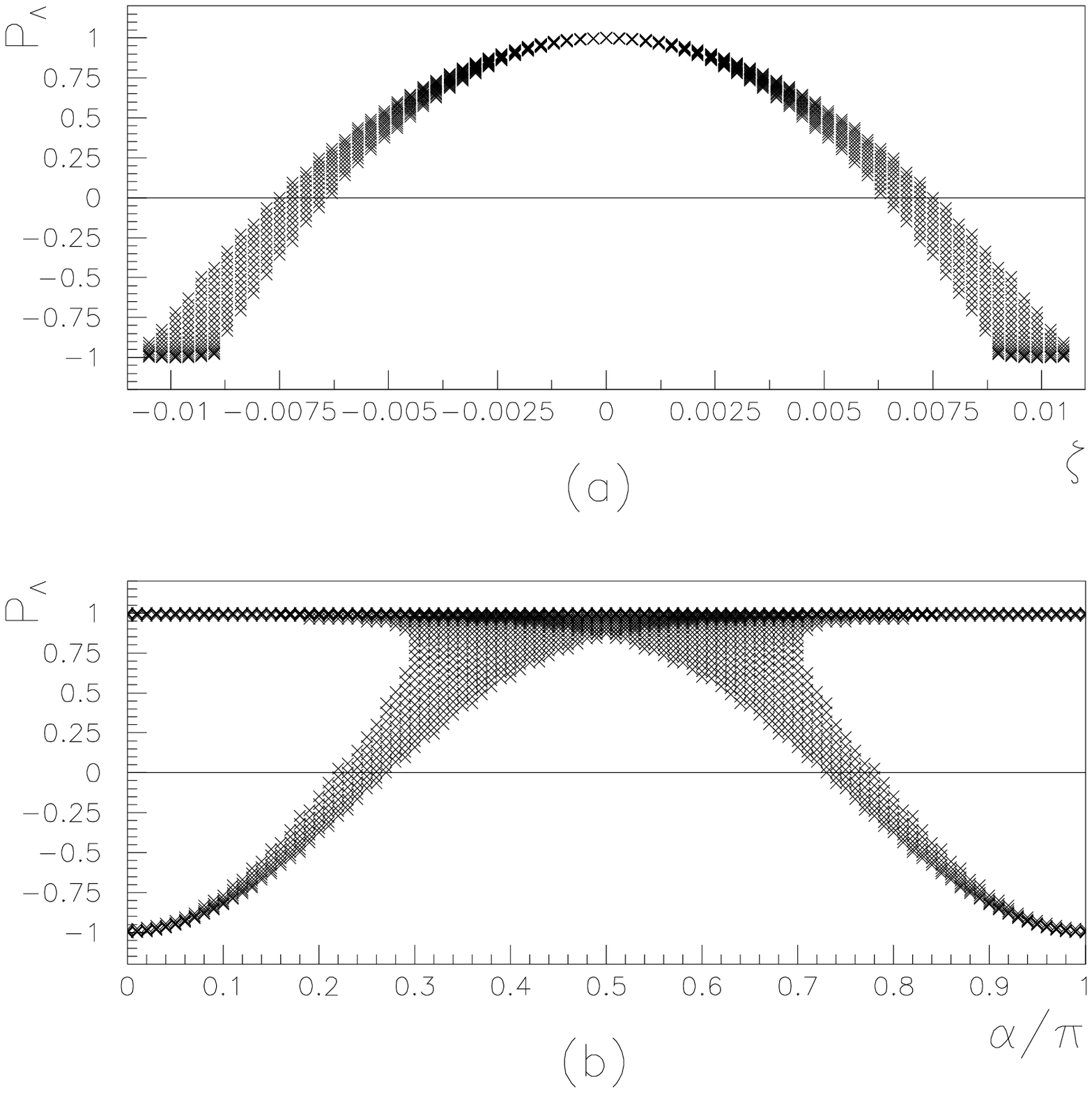}
\caption{ Dependence of $P_\Lambda$ on (a) $\zeta$,  
and on (b) $\alpha$. Here we imposed the present inclusive 
${\cal BR}(B \to  X_s \gamma)$ constraints.}
\label{fig3}
\end{figure}

\begin{figure}[ht]
\hspace*{-1.0 truein}
\psfig{figure=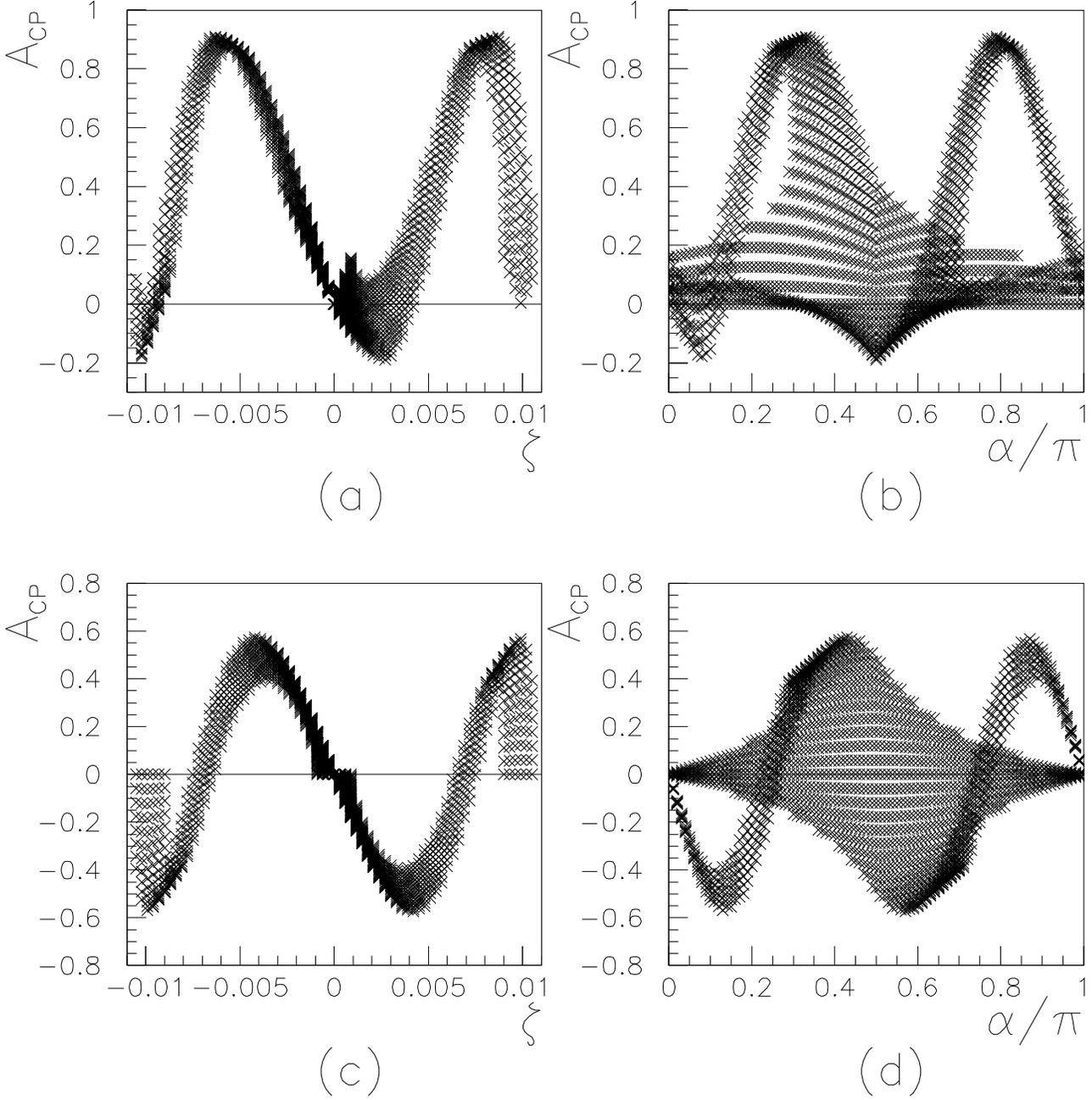}
\caption{ Dependence of $A_{CP}$ on $\zeta$, 
and on $\alpha$ is shown in (a) and (b), respectively,
for $B_d \to M^0 \gamma$ decay; and 
in (c) and (d) ,respectively,
for $B_s \to M^0 \gamma$ decay.
Here we imposed the present inclusive ${\cal BR}(B \to  X_s \gamma)$ 
constraints.}
\label{fig4}
\end{figure}

\begin{figure}[ht]
\hspace*{-1.0 truein}
\psfig{figure=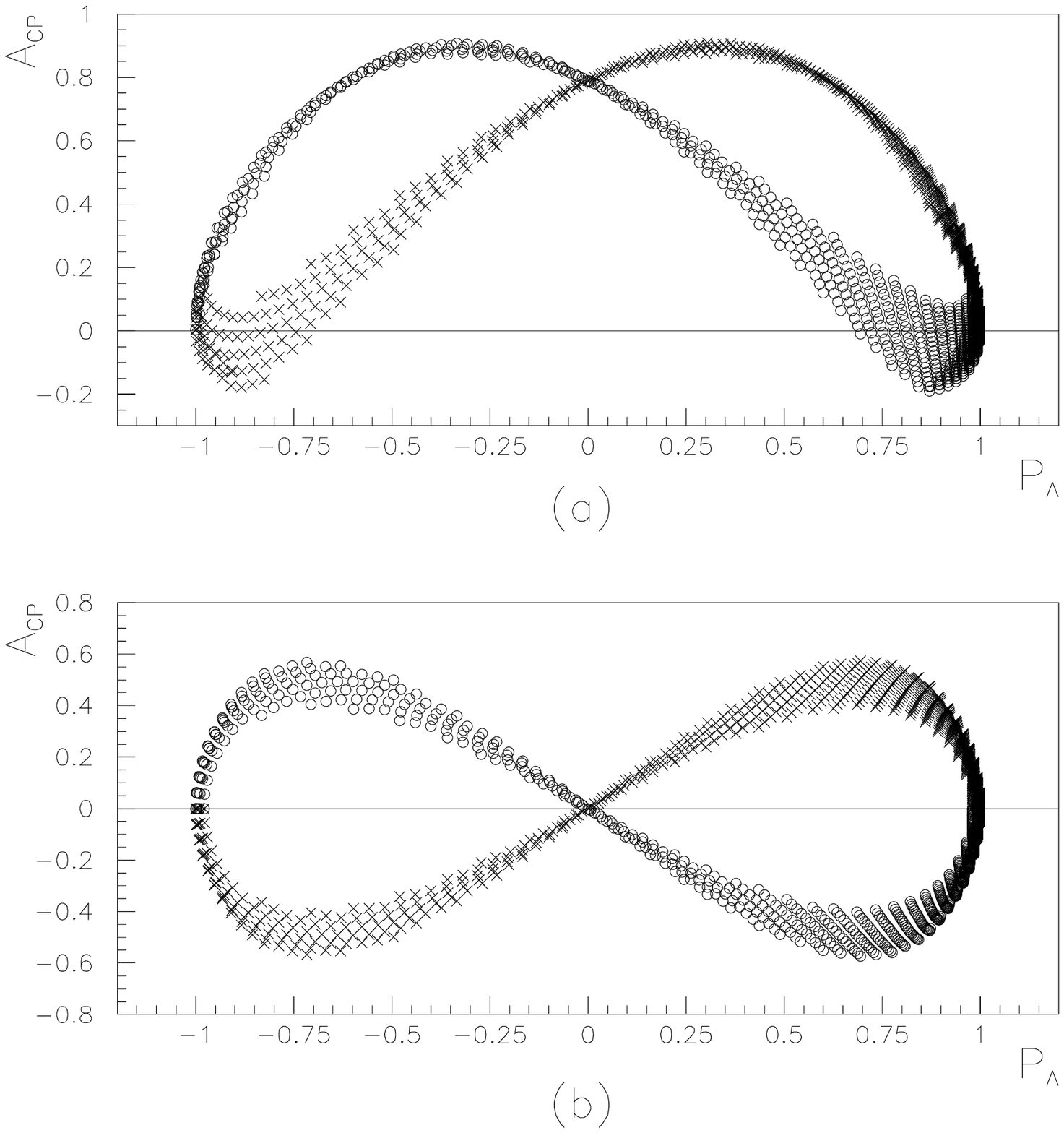}
\caption{Correlations between $A_{CP}$ and $P_\Lambda$:
(a) and (b) correspond to $B_d \to M^0 \gamma$ 
and $B_s \to M^0 \gamma$ decays, respectively.
Here we imposed the present inclusive ${\cal BR}(B \to  X_s \gamma)$ constraints.}
\label{fig5}
\end{figure}

\begin{figure}[ht]
\hspace*{-1.0 truein}
\psfig{figure=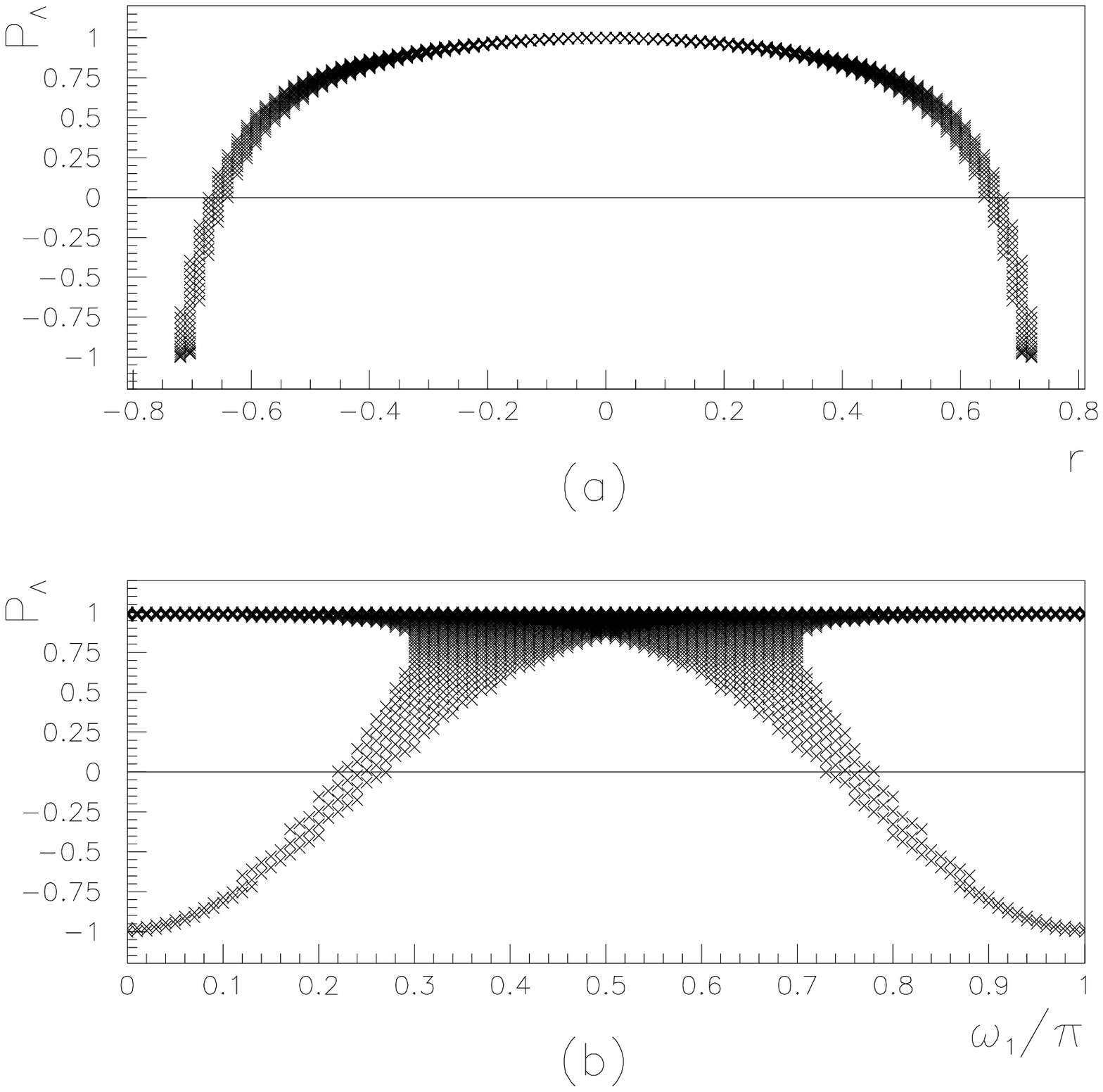}
\caption{ Dependences of $P_\Lambda$ on (a) $r$,  
and on (b) $\omega_1$. Here we imposed the present inclusive 
${\cal BR}(B \to  X_s \gamma)$ constraints and fix $M_2 = 1.6$~TeV and
$M_H = 12$~TeV.}
\label{fig6}
\end{figure}

\begin{figure}[ht]
\hspace*{-1.0 truein}
\psfig{figure=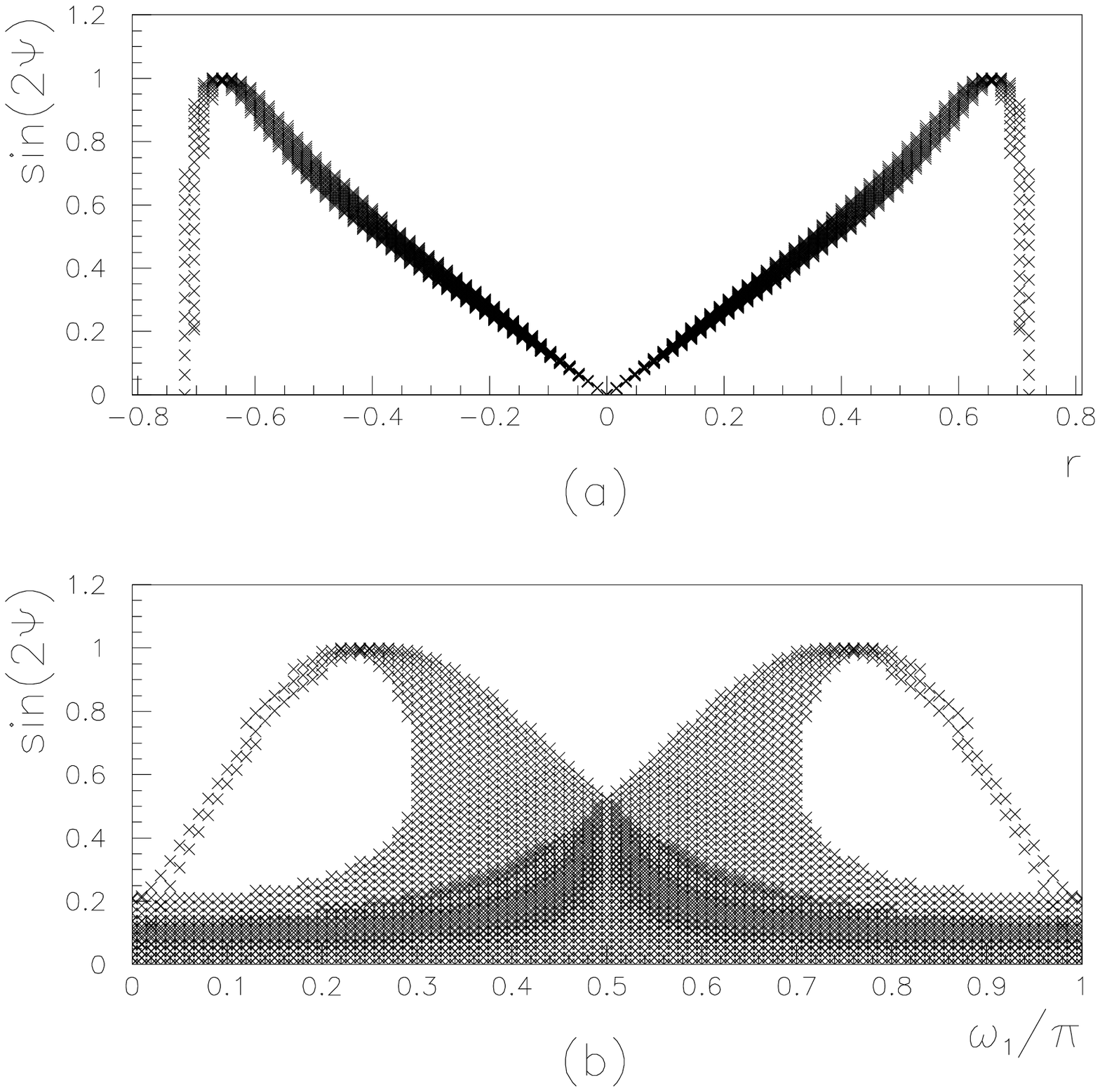}
\caption{ Dependences of $\sin (2 \Psi)$, the maximum value of $A_{CP}$, on $r$, 
and on $\omega_1$ are shown in (a) and (b), respectively,
for $B_{d,s} \to M^0 \gamma$ decay. 
Here we imposed the present inclusive ${\cal BR}(B \to  X_s \gamma)$ 
constraints and fix $M_2 = 1.6$~TeV and $M_H = 12$~TeV.}
\label{fig7}
\end{figure}

\end{document}